# Base-Rate Fallacy Redux and a Deep Dive Review in Cybersecurity

Robert F. Erbacher
U.S. Army Research Laboratory
Robert.F.Erbacher.civ@army.mil

**This paper examines the current state of the science underlying cybersecurity research with an emphasis on the non-signature-based intrusion detection domain. First, the paper re-examines the base-rate fallacy originally published by Axelsson, putting the impact of false positives into context. Given the relative high numbers of false positives, the paper argues for deeper analysis of false positives, akin to the analysis that true positives are treated to. The second section of the paper examines the metrics being used to analyze non-signature intrusion detection techniques, the current status quo of employed metrics, and the impact of the status quo on scientific advancement. Finally, the paper analyzes the use of online attack graphs and their applicability, especially in scenarios of constrained environments, such as Internet of Things devices. The use of offline attack graphs in such constrained environments is also examined. In essence, a deep dive review identified multiple areas throughout the field in which the effectiveness and validity of the scientific method can be greatly improved, e.g., through removal of logical fallacies.**

In 2000, Stefan Axelsson published his paper on "The base-rate fallacy and the difficulty of intrusion detection" [1]. In that paper, he argued that effective intrusion detection necessitates a low false alarm rate *(FAR)* in conjunction with a high true positive rate. Intrusion detection in this scenario specifically relates to non-signature-based intrusion detection, such as anomaly detection [14]. This clearly remains the case, however, there doesn't seem to be a clear understanding as to what a low false alarm rate actually entails; the base-rate fallacy Axelsson warned against remains in effect. This paper puts false alarm rates into context and examines a deep dive in cybersecurity, focusing on improving the scientific foundations of cybersecurity through examining the broader implications of logical fallacies in cybersecurity.

When Axelsson published his paper, data mining techniques were all the rage for intrusion detection but none proved viable due to not being able to achieve a false alarm rate below $10^{-3}$ [13]. For reference, Axelsson specifically defines false alarm rate to be equivalent to false positive rate *(FPR)*. Currently, machine learning techniques are, again, all the rage, however, the false alarm rates are only nominally better with most still in the range of $10^{-2}$ and the rare exception achieving $10^{-3}$ [22]. What does this mean in the context of cybersecurity and is this sufficient?

## Base-Rate Fallacy in Context

To understand the false alarm rate, we must put this into context. Specifically, the fact that each false positive must be examined by an analyst and given the raw volume of network data even a small false alarm rate will result in a huge number of actual false positives. Understanding this scale is fundamental to understanding the implications of the false alarm rate.

First, we must be able to project the number of expected false positives in a data set, based on the number of samples in the set. The false alarm rate, defined as the false positive rate, gives us the number of false positives we can expect given the number of true negatives. For projection, we will know the approximate sample size but not the number of true negatives. Thus, false positive rate may not be accurate. Preferably, we would use *FP%*, specified as $\frac{FP}{(TP+TN+FP+FN)} * 100\%$, since it gives us the number of false positives to expect given the total number of samples. With the variables defined as true positive *(TP)*, true negative *(TN)*, false positive *(FP)*, and false negative *(FN)*.

In a realistic data set, since the number of actual positives is small in comparison to the number of actual negatives, the false positive rate will be close but not equal to FP%. Care must be taken to use and interpret the metrics correctly since data sets for research testing are typically skewed with artificially high numbers of true positives; e.g., the test data in [35] has 40% true positives. Depending on the distribution of data in the test data versus the expected distribution in the real data will determine which metric will be more accurate for projection; thus, it makes sense to have a data set specifically for computing the projection of false positives.

Second, we need to consider the data source. When considering the base-rate fallacy we are typically analyzing network traffic data at the packet level; network traffic data analyzed by sessions or other sources of data are less problematic but should still be

Table 1: Analysis of analyst requirements for typical false positive percentages.

|    | Label        | Source/Equation     | Units           | Value                                |
|----|--------------|---------------------|-----------------|--------------------------------------|
| 1  | $BpD$        | [32]                | Bytes/Day       | 2500000000000.00 ($2.5 \times 10^{12}$) |
| 2  | $BpH$        | $BpD/24$            | Bytes/Hour      | 104,166,666,666.67                   |
| 3  | $BpP$        | [10]                | Bytes/Packet    | 870.607188                           |
| 4  | $PpH$        | $BpH/BpP$           | Packets/Hour    | 119,648,296.1574974                  |
| 5  | $FP\%$       | [22]                | % FP            | 0.37 *                               |
| 6  | $FPpH$       | $PpH * FP\%/100$    | FP/Hour         | 442,698.6957827404                   |
| 7  | $FPpA_nH$    | [8]                 | FP/Analyst/Hour | 12                                   |
| 8  | $A_npH$      | $FPpH/FPpA_nH$      | Analysts        | ≈36,892                              |
| 9  | $Shifts$     | [38]                | # Shifts        | 4                                    |
| 10 | Total Staff  | $A_npH * Shifts$    | Analysts        | ≈147,568 ($10^5$)                    |

* This is close to the best (selective) % FP reported by Lee et al. of .31% [13].

analyzed for the impact of the false alarm rate. Here, we focus on the ramifications of the false alarm rate with the typical packet-based approach, as seen in Table 1.

The data volume specified in [32] is 2.5 TB inbound per day. While this is from 2011, it does provide a definitively specified volume of inbound traffic. As our goal here is to show the high impact of what is currently considered low false alarm rates, the fact that data traffic has only increased from this data point strengthens the argument; while clearly rounded this definitive specification is less fuzzy than attempting to approximate actual network traffic usage from broadly specified network capacities.

Further, the alert analysis rate is identified in [8] as being 12/hr. We will assume 870.607188 bytes/packet, as derived in [10]; obviously this will vary but such variations won't significantly impact the scale of the problem.

Thus, as seen in Table 1, using the typical false alarm rates available today, perhaps even unrealistically optimistic false alarm rates and data volumes from 2011 necessitates that on average over 37K analysts need to be on staff each shift. Assuming 4 shifts [38], this would necessitate 147K analysts be on staff. Keep in mind that it is the exponent of the values in the table that are important and not the exact number of staff. Even if some numbers shift a bit (bytes per packet, # shifts, etc.), the exponents, and thus the scale of the problem, won't shift appreciably. Given the data volume used is from 2011, the actual scale of the problem is only likely to be worse.

Further, this is for just one tool. Analysts will often use dozens, if not 100s of tools; a study by Silva et al. [27] observed 75 distinct tools being used in a cyber-exercise by 11 subjects. Even if a tool ends up being the epitome of all tools and no others are needed, false positives need to decrease significantly before such tools are even remotely viable. Note that in the best case the $FP\%$ have only decreased by $10^1$ in the nearly twenty years since Axelsson's original base-rate fallacy paper was written, yet the volume of data has increased by significantly more than $10^1$. The issues that caused Axelsson to write his initial paper remain issues today but his paper has fallen out of the collective memory.

As an alternative mechanism for evaluating the volume of false alarms being generated by a tool, consider: "A survey by FireEye polled C-level security executives at large enterprises worldwide and found that 37 percent of respondents receive more than 10,000 alerts each month. Of those alerts, 52 percent were false positives, and 64 percent were redundant alerts." [7] This amounts to about 14 total alerts per hour, which further exemplifies and validates the unacceptability of techniques currently being reported in research results. Comparing the results in line 6 of Table 1, instead of a total of 14 alerts per hour, the techniques under discussion are generating 442K alerts just for false positives. While the author's specification indicates this 14 alerts/hour is a minimum, the scale and coverage of all alerts delineates the difference between actual and needed scale of alerts for non-signature-based detection techniques.

## Consideration of False Positives

Too often analysis focuses on metrics with only limited analysis of what attacks are actually detected or not. For instance, are we not detecting the most significant attacks that are most likely to succeed in the real world? Similarly, there is typically no analysis of false positives. Typically, they are reported without comment. We need less reporting of metrics and more justification of them. There has been no identifiable improvement in the generation of false positives but the field keeps publishing without additional techniques to handle the false positives. This simply isn't leading to viable solutions.

Given the significance of false positives we cannot just report their metrics and assume they are "fine". This can't continue to be the case; they must be analyzed – both their nature and their impact. The argument has been made that many, if not most, of the false positives can essentially be relegated to trivial accepts or trivial rejects, e.g., through triage. Thus, the argument is that the analyst can quickly skip over such alerts when scanning through the list of alerts. However, this is never validated. Given the difficulty in substantially reducing false positives and their critical impact to the viability of non-signature-based detection techniques for cyber, we must begin considering and analyzing false positives in a more complete fashion. As mentioned previously, we need data sets designed to allow projection of the number of false positives. Similarly, we need data sets to allow complete analysis of the nature of false positives. Given the significance of the false positives in non-signature intrusion detection, taking these actions will improve the scientific foundation of intrusion detection.

In essence, during research experimentation, we must analyze false positives as we would true positives, with the goal of understanding their nature and characteristics. As we analyze true positives to determine what classes of attacks are detected and at what rates, we must analyze false positives to understand their nature, similarity, and ease with which they can be subsequently purged.

While not currently done, the potential for pre- and post- processing techniques to eliminate a majority of the false positives could assist non-signature-based techniques in gaining viability. At the pre-processing level this could include eliminating packets verified to be valid components of a session, such as a VPN connection. At the post-processing level this could include eliminating replicated alerts. The potential for pre- and post- processing needs to be researched independently. Further, general approaches that reduce false positives must be examined if non-signature-based approaches are to be made viable.

## Extensions to Metrics

The previous section identified issues in published works using standard metrics. It is important to understand the metrics and their relevance to cybersecurity. Many of the metrics we use for analyzing cybersecurity techniques come from information retrieval [21] and from a purely information theoretic point of view are valid. However, a deep dive examination relative to the specific context of cybersecurity shows that not all of the formulas currently being used are relevant; science can be improved by validating relevance. For instance, consider accuracy:

$$Accuracy = \frac{TP + TN}{FP + TP + FN + TN}$$

Accuracy is seeing frequent usage and conceptually it makes sense. However, formulaically, its reliance on TN in the numerator, giving it equal weight to TP, is invalid in cybersecurity due to the high volume of $TN$s relative to $TP$s and their nonexistent value. It's FPs that are a concern, in volume, while $TN$s are completely irrelevant. This is, in essence, an additional example of the base-rate fallacy in operation.

Another way to interpret this is through value/cost. A TN literally has zero value. If a lost packet is important it will be resent and a received packet that isn't part of an attack has zero value as far as cybersecurity is concerned. A TP on the other hand is an attack or part of an attack and we have to associate with it the average cost of a successful attack, currently $3.92M [36]. When we have an accuracy value not of 100%, we have no idea what number of the incorrect labels are zero-value $TN$s or high-value $TP$s. The best practice necessitates the assumption that all of the incorrect labels be associated with $TP$s, thus incorporating $TN$s doesn't follow logically.

Essentially, the issue is a deviation between a theoretic assessment of the technique and a real-world assessment. From a purely theoretical point of view, accuracy appears to be a relevant and completely applicable metric. From the real world, cybersecurity point of view, accuracy is completely irrelevant. The problem being that accuracy assumes all samples are of equal weight. Since that isn't the case, accuracy cannot be used.

### Looking Beyond Accuracy: False Alarm Rate

Beyond accuracy, scientific foundations can be improved through improved understanding of the metrics they are using, what their definitions are, and their relevance to the task at hand. Exemplifying the problems of the domain, there appear to be an unlimited number of definitions for false alarm rate. To be clear, authors are definitively specifying different formulations for the metric "false alarm rate", for example:

- $FP/(TN + FP)$ [22][15][33], i.e., equivalent to $FPR$

- $A/¬I$, alarm/nonintrusive behavior [1], i.e., specified by the author to be $FPR$ though the specific formula isn't given.

- "$FAR$: the number of NORMAL cases that are detected as intrusions in a specific category divided by the total number of NORMAL cases." [11], i.e., also equivalent to $FPR$ though not formally specified mathematically.

- Undefined [5]. In essence, even back as far as 2002 authors were not defining false alarm rate definitively, demonstrating the duration of the problem.

- $FP/(TP + FP)$ [23], i.e., equivalent to false discovery rate *(FDR)*

- $FN/(FN + TP)$ [4] [29][12], i.e., equivalent to false negative rate *(FNR)*

- $(FPR + FNR)/2$ [19]

- "it is the addition of $FNR$ and $FPR$" [18], as with accuracy, this metric and the previous add unequal classes, which is ineffective

Note that the multiple conflicting definitions aren't simply a result of one-offs as there are many papers with *alternative* definitions and many papers with the same *alternative* definition. Many of these *alternative* definitions are provably invalid. Other than the fact that there are multiple conflicting definitions, this list has five specific problems:

First, note that [22] defines false alarm rate using the formula for false positive rate but in the same table uses $FPR$ without a defined formula; they do, unfortunately, have different values. Further, the text associates false alarm rate with the value associated with $FPR$ in the table. I am operating under the assumption that $FPR$ in this paper is intended to be what I am defining as $FP\%$ but the lack of the underlying data prevents validation; the definitive typos substantiates the fact that archival papers should include sufficient information to fully validate/replicate their results; in this case the metrics themselves.

Second, $FN$ are not alarms be definition. Thus the last three equations which are dependent on FN in the numerator are simply not valid equations for a false alarm rate; they should have been true alarms. This doesn't mean such formulations don't have value, they simply aren't validly termed false alarm rates. This failure exemplifies the need to reassess the metrics currently being employed and ascertain their validity; the fact that I was able to identify multiple uses of these metrics reinforces the extent of the problem; i.e., this isn't a single isolated case.

Third, in the case where the false alarm rate is defined as $FN/(FN + TP)$, note they specify $FN$ and not $FP$ in the numerator. There is insufficient information to validate where the error is, is it just a typo in the equation or are the numerical results wrong as well. Having false negatives in the numerator does not make sense from a language perspective since they are not alarms. This could have simply been a typo, which since it has been identified in multiple papers, could be the mistake simply being carried on by other researchers. With the authors only reporting the resulting metrics and not the actual $TP$ and $FP$ rates it is impossible to validate. This exemplifies a greater failing in terms of providing sufficient information in papers for replication, validation, and longevity.

Fourth, the lack of definitive specification in cases leads to the inability to validate the results. We've already identified several cases where invalid metrics are being used, possibly the result of typos in some cases, and we've identified four definitive different specifications for false alarm rate. Without a definitive specification, such as the paper that I had to label the formula as undefined, which are we using, if any?

Finally, note that some specifications simply use English to specify the formula and do not provide a formal mathematical equation. This can lead to problems in interpretation, especially for non-native speakers, and can lead to issues with differing specification.

## Looking Further: Detection Rate

In addition to false alarm rate, the second major metric used by researchers for intrusion detection is detection rate *(DR)*, which suffers from the same fallacies as false alarm rate. This is obviously critical since it gets at the heart of the purpose of intrusion detection – the ability to detect attacks. However, as with false alarm rate, detection rate itself is not a well-established metric. As with false alarm rate, this results in multiple conflicting definitions:

- $TPR$ [18][33][22]; this definition matches the majority of uses

- $P(A/I)$ [1], i.e., specified by the author to be $TPR$ though the specific formula isn't given. Similarly, TPR is used but not defined in [6]

- $TP/(TP + FP)$ [15], precision or positive predictive value *(PPV)*

- $TN/(TN + FP)$ [12], true negative rate *(TNR)*. This is just wrong. Another potential typo that can't be validated.

- "DR means the rate of data that are correctly classified into normal and intrusion classes". [16] Note that this would imply a definition equivalent to accuracy with the inherent flaws previously described related to accuracy but since the authors didn't provide a mathematical specification there is some uncertainty about the incorporation of "normal classes", i.e., TN, implies a formula more presentative of accuracy as opposed to TPR. However, these authors also use accuracy, again without a formal specification, but with different results from DR. This is yet another example of the lack of complete specification leading to the inability to interpret or validate presented results.

- "that represents the percentage of correctly detected intrusions, and false positive, that represents the percentage of normal connections that are incorrectly classified as anomalous." [26]

The same issues identified for false alarm rate exist with this list.

## Improving Science through Metric Reduction

The state of metrics in the cyber intrusion detection domain is exemplified by the fallacy that the results are meaningful and advance the science. In Umer et al. [30] three tables of results are presented comparing the effectiveness of different techniques. In each table, the "Performance Measure" is a column in the table, preceding the result column. While result "values" are then presented, there is simply no meaningful way to compare the techniques to each other given the diversity of metrics being applied. The "Performance Measures" referenced in these tables include:

- Anomaly type detection
- Area under ROC curve
- Botnet detection
- Brute force Detection
- Class Prediction
- Recall
- Correctness rate
- Custom
- Detection Accuracy
- Detection rate
- DF Rate
- F1-measure
- F1-score
- Comparison
- ROC Curve
- Graph Generation
- No. of TCP scans detected
- Purity measure
- True Positive
- True Positive Rate

This is further exemplified in [35] which lists the primary evaluation methods for 19 papers with the following metrics and distributions: FPR (5), detection rate (3), accuracy (7), true positive rate (1), true positive (2), false negative (2), precision and recall (2), negative predictive value (1), error rate (1), N/A (1). There is clearly overlap, however, it is insufficient for scientific rigor. The inconsistency demonstrates the current impossibility of comparing techniques over time.

This is simply not viable for the advancement of the field. You cannot have 20 different result metrics and be able to determine if a technique is an improvement or not. This is in addition to the fact that we have demonstrated some of these metrics having many different specifications themselves. Authors simply are not citing where their metrics are coming from. This makes it impossible to determine the extent, source, motive, etc. spawning the errors.

A final issue is demonstrated in Mishra et al. [18] in which the authors directly compare – through tables and text – detection rates and false alarm rates between papers in which these values use different formulations. Further, the authors do not call out the different formulations or recompute the metric values. A reader would assume that the results would be computed using the same formula but they are not. The implication is that the author didn't realize they were comparing different metrics since they were all labeled identically.

Specifically, [18] directly compares the detection rates presented in [15], [12], and [33], which as shown above are each computed differently. Further, [18] directly compares the false alarm rate of [15] and [33] with [12], which are also (as demonstrated above) computed differently.

Further, the detection rate and false alarm rate that the authors of [18] present for the metrics in [12] do not come from [12] and there is no explanation as to where these values derive from. Specifically, instead of simply reporting the detection rate presented in [12] the authors present the *overall detection rate* but do not explain the derivation of the metric. I was not able to derive the values presented in [18] from the values presented in [12].

In essence, they are comparing papers based on metrics with reported differences in their specification, without specifying any deviations from the reported metrics. This, combined with the deviations in values between the original paper and the citing paper makes it impossible for a reader, even with close attention and detailed analysis, to comprehend what exactly is going on or how to use the results.

## Defining the Metrics

There is clearly a need to get researchers on the same page as far as metrics are concerned. This includes using the same definitions for false alarm rate and detection rate but also including sufficient information for validation. There are already existing sources documenting well-established metrics. This includes Wikipedia [37] which documents the metrics from information retrieval and [21] which documents many well-established metrics along with analysis of their biases. Thus, this paper doesn't delve into all the available metrics or analysis of them but rather

Table 2: Proposed metrics to be used, their specification, and description.

| Name | Label | Formal Specification | Description/Explanation |
|---|---|---|---|
| True Positives | TP | | Empirical result value. True positives identifies the set of actual positive samples reported as positives |
| False Positives | FP | | Empirical result value. False positives identifies the set of actual negative samples reported as positives |
| Sample Size | N | | Sample size of the test |
| Database | DB | | Sufficient information must be presented or referenced about the data source to ensure accurate validation and replication. This could reference a data set or specifically identify modifications to a data set |
| Precision | PPV | $\frac{TP}{TP + FP}$ | The fraction of returned positive results that are actually positive, i.e., the likelihood that a positive test result is actually a positive |
| Recall | TPR | $\frac{TP}{TP + FN}$ | The fraction of actual positives that are identified as positive, i.e., the fraction of identified positives |
| F-Measure | $F_1$ | $2 \times \frac{precision \times recall}{precision + recall}$ | The harmonic mean of precision and recall. Provides precision and recall as a single metric for more effective direct comparison between techniques |
| False Positive Rate | FPR | $\frac{FP}{FP + TN}$ | The fraction of actual negatives that are improperly identified as positive |
| False Positive Percentage | FP% | $\frac{FP}{FP + TP + FN + TN} \times 100\%$ | The percentage of all samples improperly identified as positive |
| Diagnostic Odds Ratio | DOR | $\frac{TP \times TN}{FP \times FN}$ | "A test with high specificity and sensitivity with low rate of false positives and false negatives has high DOR." [28] |
| False Negatives | FN | | Empirical result value. False negatives identifies the set of actual positive samples reported as negatives |

recommends metrics for consistency, depth of analysis, validation, and appropriateness.

It is further necessary to use the same terminology consistently as metrics have different formal terminology for the same mathematical specification. For instance, recall is identical to TPR and sensitivity with both recall and sensitivity being used broadly. The state of field demonstrated in this paper necessitates more consistency.

While many metrics have been used, including the aforementioned, other common metrics are F-Measure, Recall and Precision, as well as Sensitivity and Specificity. The proposed metrics to be included are in Table 2.

True Positives, False Positives, Sample Size, and Database are included for validation and replication. This may seem pedantic but the current state necessitates the detail. Precision, Recall, and F-Measure are the standard metrics currently used to assess cybersecurity techniques. As discussed, so many metrics have been used that it is impossible to provide compatibility with all of them so we limit this to the most common metrics. Additional metrics can always be derived, the goal is to provide the most common metrics for rapid assimilation.

False positive rate and false positive percentage are included to emphasize the importance of considering false positives as a critical result. Diagnostic Odds Ratio [28] is included as a potentially more effective metric and is proposed as a new standard to be included. False Negatives is included for completeness as it is needed to compute already mentioned metrics. Given the lack of consistency in metric definition, the goal is to be complete to avoid any further confusion. The terms (metrics?) *false alarm rate* and *detection rate* were not included as they've become too compromised to have

| | Blocklist | Allowlist General System | Allowlist Constrained System |
|---|---|---|---|
| **Table 3: Analysis of Scale and Impact of Blocklisting versus Allowlisting.** | | | |
| Variables | $TP = K$ $FP = C_k$ $TN = C_a + K + C_u - A - C_k$ $FN = A - K$ | $TP = C_a + K + C_u - A - C_k$ $FP = A - K$ $TN = K$ $FN = C_k$ | $TP = C_a - A - C_k$ $FP = A - K$ $TN = K$ $FN = C_k$ |
| PPCR | $\dfrac{C_k + K}{C_a + C_u + K}$ | $\dfrac{C_a + C_u - C_k}{C_a + C_u + K}$ | $\dfrac{C_a - C_k - K}{C_a}$ |
| Recall | $\dfrac{K}{A}$ | $\dfrac{C_a + C_u + K - A - C_k}{C_a + C_u + K - A}$ | $\dfrac{C_a - A - C_k}{C_a - A}$ |

any meaning and given their imprecision should not be used.

A couple of notes about the Diagnostic Odds Ration. First, its inclusion of true negatives in the numerator is less problematic than accuracy due to the use of multiplication rather than addition, but it is also not attempting to represent the same concept as accuracy and as an overall metric is effective. Second, the nature of the metric means it is not as effective at representing effectiveness of techniques across data sets. However, this is a good point as too often comparing metrics generated between different data sets should not be compared due to deviations in signal to noise ratio, attack coverage, etc.

Finally, publications must formally specify the mathematical formulations for each metric within the paper. This can be done by specifically including the definitions of the metrics or citing a reliable source for the specifications and specifically stating the definitions came from that source. The examples in this paper have demonstrated the need for these measures.

## Extensions to Attack Graphs

A final approach towards improving the scientific foundations of cybersecurity is through reassessing baselines to ensure the most effect approach/techniques are being employed; e.g., don't assume the baselines do not need reassessment. Specifically, attack graphs, originally termed attack trees by Schneier [25], came about as a result of the need to identify the threats and vulnerabilities relevant to specific systems and the difficulty in securing said systems. Attack graphs were later adapted to online analysis by Noel et al. [20].

The systems being considered at the time attack graphs were first being developed were all full-fledged and fully capable systems, with multi-user systems the primary target of concern. Since that time, the advancement and deployment of embedded systems have exploded with the likes of IoT devices, UAVs, smart cars, etc.

Thus, Capobianco et al. [3] updated online attack graphs for modern needs of intrusion detection, including embedded systems. Similarly, earlier work by Vasilevskaya et al. [31] updated offline attack graphs for embedded systems. This is primarily where we run into problems. First, it is important to keep in mind that the idea of online attack graphs is for detection and not protection. Second, both online and offline attack graphs focus solely on known attacks and vulnerabilities, they provide no mechanism for unknown attacks. However, it is the unknown attacks, the zero-day attacks, that are the greatest threat. This results from the fundamental nature of attack graphs as essentially attempting to blocklist known bad behavior.

## General Purpose Systems

The focus on blocklisting results from the historical impossibility of allowlisting; even though allowlisting is known to be more secure. The impossibility results from the sheer volume of valid activities in a general-purpose computing device; which is even further exacerbated in multi-user systems.

This is demonstrated in Table 3 in which blocklisting is compared to allowlisting in general use systems. In this table, $A$ is the set of anomalous actions and action sequences (just referred to as actions from here on), $K$ is the set of *known* anomalous actions, $C_k$ is the set of innocuous actions matched by known malicious signatures, $C_u$ is the set of actions that are unused, unneeded, or deemed unsafe for the controlled and constrained embedded system, and $C_a$ is the set of actions deemed to be acceptable for the controlled and constrained embedded system.

The scale of the entire set of valid actions and action s sequences is shown in the denominator of predicted positive condition rate (PPCR); with PPCR formulated as $(tp + fp)/(tp + fp + tn + fn)$. The numerator of PPCR represents the total positive responses generated by the system. For a general use system, the set of valid actions and action sequences is equivalent for allowlists and blocklists but the set of generated positives is significantly higher. Note that PPCR provides the percentage of the valid actions and action sequences resulting in positive assertions.

However, note that blocklisting cannot detect unknown attacks. This is represented in the table through the Recall metric, also termed the true positive rate, which provides the percentage of the total set of relevant events that are actually successfully identified. For blocklisting this is attempting to detect all attacks. For allopwlisting this is attempting to identify the set of valid commands.

## Constrained Systems

The situation changes, however, when we consider embedded systems. Specifically, with embedded systems we are considering systems in which we do not want the system treated as a general-purpose system but rather where the system should be tightly controlled and constrained. Such systems will typically still have general purpose CPUs and fully functional communication systems, albeit with lower performance capability.

Consider a UAV for instance, it may be built with general purpose capabilities but in general use, especially when in flight, only very limited capabilities are actually needed or should actually be allowed. In addition to the extent to which actual functionality should be limited, there is an increase in the control over how the functionality should be used or accessed. Since we are not dealing with general-purpose systems, we can enable extensive controls over the use of any and all functionality. With such a controlled and constrained environment, the system can purge anything not precisely matching what is required.

This change in the baseline for embedded system is the impetus for employing allowlisting as opposed to blocklisting. Blocklisting simply cannot detect unknown attacks so it's less effective than allowlisting. Blocklisting can still have benefits in identifying weaknesses in the allowlisting but it must serve a secondary role to allowlisting, .i.e., in an offline mode, whereas allowlisting to this point has not been seriously investigated. In essence, the argument for using blocklisting fails due to the reduced scale of valid actions and action sequences and the greater effectiveness of allowlisting.

This is demonstrated in Table 3 through the constrained system column. The scale of blocklisting remains the same remains the same since we are still typically dealing with general purpose components. The scale of allowlisting, however, clearly becomes significantly reduced; to the point where the scale is smaller than with blocklisting, likely significantly. This is all while being more effective at preventing attacks than blocklisting.

Whereas the blocklisting being employed is specifically attack graphs, we propose the idea of validity graphs for the implementation of allowlisting. We are in no way arguing that switching to validity graphs will be easy, they still require research. However, the goal will be to identify minimal action graphs that allow needed functionality; everything else would be considered an attack. The constrained environment should result in a significantly reduced number of actions and action sequences needing validation. Further, validity graphs can disallow not just attacks but invalid commands. For instance, in the UAV example, setting pitch (-1) could be deemed invalid as it would eventually lead to a crash. Since we can precisely control how actions are specified, we can ensure that this action is not specified in isolation. Additional functionality could always be made accessible though techniques along the lines of port knocking [34].

Thus, there is a fallacy in assuming that the baseline is still valid and "jumping on the bandwagon" even when the status quo has changed substantially. This again leads to bad science. In this case, there is a valid need to consider alternatives to attack graphs. Attack graphs still have validity in such constrained environments for network design, even in conjunction with validity graphs, but validity graphs should be examined for online analysis.

## Conclusion

It is important to understand the extent and number of papers published in major venues which exhibit problems. The demonstrated problems include inadequately addressing false positives, inconsistent metrics, and not reassessing baselines in relation to allowlisting versus blocklisting. This indicates a ubiquitous problem with evaluation in the field, with it being nearly impossible to accurately assess or interpret the value of refereed techniques. This is reminiscent of the replication crisis in psychology [2]; with recent arguments of a validation crisis [24]. Within cyber security itself, there has been a discussion of the need for more accurate application of the scientific method, rather than the lip service it appears to be getting [9]. However, this prior work was an abstract analysis of the scientific method (philosophy of science [17]) within cybersecurity. This paper provides specifics as to how the science can be improved, particularly in terms of rigor. This concrete analysis of the science underlying cyber security provides context as to how cyber security advancement can be improved through better science with a greater impetus for re-evaluating the rigor of the scientific method than the philosophical discussion alone.

Authors must investigate the papers they are citing and not take the metrics and values at face value. Papers are being written too formulaically en masse, with too little in-depth consideration being given to real meaning and impact in the desire to get papers published in quantity. Intrinsically, we must use consistent metrics to validate that science is advancing. Given the errors described in this paper that is not a given. More complete in-depth analysis is needed, not just reporting the metrics, to truly understand the value, or lack thereof, of reported research.

It is important to note that this is not a complete survey, yet the work clearly demonstrates significant problems with publications within the intrusion detection research community. Many of the examples came from the first five pages of Google scholar simply using "intrusion detection techniques" as the search term and limited to the 2017- timeframe -accessed in the fourth quarter of 2019- demonstrating that these are significant publications, likely to be cited, and not bottom tier. A deeper dive into existing publications is warranted; it must not be construed that this paper documents all problems underlying the scientific method in the cited papers.

Further, this paper does not examine work outside of the intrusion detection subfield but researchers at large should consider the implications of this work more broadly. Authors must not assume that the papers they are referencing are correct and without need of validation; the scientific method necessitates greater rigor. Improving the status quo necessitates publications that are more replicable and validatable. The safest approach would be to assume that referenced papers are not scientifically valid and ensure sufficient validation and replicability independently. Finally, the status quo must be intermittently reassessed as slow drifts in scientific merit over time are hard to detect as they are occurring but the ramifications can be seen through deep dive reviews.

## ■ REFERENCES

1. S. Axelsson, "The base-rate fallacy and the difficulty of intrusion detection," in ACM Transactions on Information and System Security, vol. 3, no. 3, pp. 186-205, August 2000.
2. C. F. Camerer, A. Dreber, F. Holzmeister, T. Ho, J. Huber, M. Johannesson, M. Kirchler, G. Nave, B. A. Nosek, T. Pfeiffer, A. Altmejd, N. Buttrick, T. Chan, Y. Chen, E. Forsell, A. Gampa, E. Heikensten, L. Hummer, T. Imai, S. Isaksson, D. Manfredi, J. Rose, E. Wagenmakers, and H. Wu, "Evaluating the replicability of social science experiments in Nature and Science between 2010 and 2015," in Natural Human Behavior, vol. 2, no. 9, pp. 637–644, 2018, doi: 10.1038/s41562-018-0399-z.
3. F. Capobianco, R. George, K. Huang, T. Jaeger, S. Krishnamurthy, Z. Qian, M. Payer, and P. Yu, "Employing attack graphs for intrusion detection," In Proceedings of the ACM New Security Paradigms Workshop (NSPW '19), New York, NY, USA, 2019, pp. 16–30, doi: 10.1145/3368860.3368862.
4. Y. Chang, W. Li, and Z. Yang, "Network intrusion detection based on random forest and support vector machine," in Proceedings of the IEEE International Conference on Computational Science & Engineering and Proceedings of the IEEE International Conference on Embedded and Ubiquitous Computing, July 2017, pp. 635–638.
5. P. Dokas, L. Ertoz, V. Kumar, A. Lazarevic, J. Srivastava, and P.N. Tan, "Data mining for network intrusion detection," In Proceedings of the NSF Workshop on Next Generation Data Mining, November 2002, pp. 21-30.
6. V. Ford, and A. Siraj, "Applications of Machine Learning in Cyber Security," in the 27th International Conference on Computer Applications in Industry and Engineering (CAINE), 2014.
7. R. Francis, "False positives still cause threat alert fatigue," in CSO Online, May 3, 2017. [Online]. Available: https://www.csoonline.com/article/3191379/false-positives-still-cause-alert-fatigue.html.
8. R. Ganesan, S. Jajodia, and H. Cam, "Optimal Scheduling of Cybersecurity Analysts for Minimizing Risk," in ACM Transactions on Intelligent Systems and Technology (TIST), vol. 8, no. 4, Article 52, February 2017, 32 pages.
9. C. Herley and P. C. van Oorschot, "Science of Security: Combining Theory and Measurement to Reflect the Observable," in IEEE Security & Privacy, vol. 16, no. 1, pp. 12-22, January/February 2018.
10. P. Jurkiewicz, G. Rzym, and P. Boryło, "How Many Mice Make an Elephant? Modelling Flow Length and Size Distribution of Internet Traffic," 2018 [Online]. Available: arXiv: 1809.03486.
11. E. Kabir, J. Hu, H. Wang, and G. Zhuo, "A novel statistical technique for intrusion detection systems," in Future Generation Computer Systems, vol. 79, pp. 303-318, 2018.
12. F. Kuang, W. Xu, and S. Zhang, "A novel hybrid KPCA and SVM with GA model for intrusion detection," in Applied Soft Computing, vol. 18, pp. 178–184, May 2014.
13. W. Lee and S. J. Stolfo, "Data mining approaches for intrusion detection," In Proceedings of the 7th conference on USENIX Security Symposium, USENIX Association, San Antonio, TX, January 1998.
14. H. Liao, C. R. Lin, Y. Lin, and K. Tung, "Intrusion detection system: A comprehensive review," in Journal of Network and Computer Applications, vol. 36, no. 1, pp. 16-24, 2013.
15. W. Lin, S. Ke, and C. Tsai, "CANN: an intrusion detection system based on combining cluster centers and nearest neighbors," in Knowledge-Based Systems, vol. 78, pp. 13-21, 2015.
16. N. Lu, S. Mabu, T. Wang, and K. Hirasawa, "An Efficient Class Association Rule-Pruning Method for Unified Intrusion Detection System using Genetic Algorithm," in IEEE Transactions on Electrical and Electronic Engineering, vol. 8, no. 2, pp. 164-172, 2013.
17. J. D. McLean, C. Herley, and P. C. Van Oorschot, "Letter to the Editor," in IEEE Security & Privacy, vol. 16, no. 3, pp. 6-10, May/June 2018.
18. P. Mishra, V. Varadharajan, U. Tupakula, and E. S. Pilli, "A detailed investigation and analysis of using machine learning techniques for intrusion detection," in IEEE Communications Surveys and Tutorials, vol. 21, no. 1, pp. 686-728, 2019.
19. N. Moustafa and J. Slay, "A hybrid feature selection for network intrusion detection systems: Central points," 2017 [Online]. Available: arXiv:1707.05505.
20. S. Noel, E. Robertson, and S. Jajodia, "Correlating intrusion events and building attack scenarios through attack graph

distances," in the 20th Annual Computer Security Applications Conference, Tucson, AZ, USA, 2004, pp. 350-359.
21. D. M. W. Powers, "Evaluation: from Precision, Recall and F-measure to ROC, Informedness, Markedness and Correlation," in Journal of Machine Learning Technologies, vol. 2, no. 1, pp. 37-63, 2011.
22. S. Prabavathy, K. Sundarakantham, and S. M. Shalinie, "Design of cognitive fog computing for intrusion detection in internet of things," in Journal of Communications and Networks, vol. 20, no. 3, pp. 291-298, 2018.
23. O. Salem, A. Makke, J. Tajer and A. Mehaoua, "Flooding attacks detection in traffic of backbone networks," in Proceedings of the IEEE 36th Conference on Local Computer Networks (LCN), 2011, pp. 441–449.
24. U. Schimmack, "The Validation Crisis in Psychology," Accessed: 11/18/2019. [Online]. Available: https://replicationindex.files.wordpress.com/2019/04/validation.crisis.v3.pdf.
25. B. Schneier, "Attack trees," in Dr. Dobb's Journal, vol. 24, no. 12, pp. 21-29, December 1999.
26. H. Sedjelmaci and M. Feham, "Novel Hybrid Intrusion Detection System for Clustered Wireless Sensor Network," in the International Journal of Network Security & Its Applications (IJNSA), vol. 3, no. 4, July 2011.
27. A. Silva, J. Mcclain, T. Reed, B. Anderson, K. Nauer, R. Abbott, and C. Forsythe, "Factors impacting performance in competitive cyber exercises," in the Proceedings of the Interservice/Interagency Training, Simulation and Education Conference, 2014.
28. A. Šimundić, "Measures of Diagnostic Accuracy: Basic Definitions," in Electronic Journal of the International Federation of Clinical Chemistry (EJIFCC) vol. 19, no. 4, pp. 203-211, 2009.
29. I. S. Thaseen and C. A. Kumar, "Intrusion detection model using fusion of chi-square feature selection and multi class SVM," in Journal of King Saud University Computer and Information Sciences, vol. 29, no. 4, pp. 462-472, 2017.
30. M. F. Umer, M. Sher, Y. Bi, "Flow-based intrusion detection: techniques and challenges," in Computer Security, vol. 70, pp. 238–254, 2017.
31. M. Vasilevskaya and S. Nadjm-Tehrani, "Quantifying risks to data assets using formal metrics in embedded system design," In Proceedings of SAFECOMP. 2015. LNCS, vol. 9337, Springer, Cham, 2015, pp. 347–361, doi: 10.1007/978-3-319-24255-2_25.
32. H. Villa and E. Varki, "Characterization of a campus internet workload," In the 27th International Conference on Computers and their Applications (CATA), March 2012, pp. 140-147.
33. H. Wang, J. Gu, and S. Wang, "An effective intrusion detection framework based on SVM with feature augmentation," Knowledge-Based Systems, vol. 136, pp. 130–139, November 2017.
34. D. Worth, "COK: Cryptographic one-time knocking," Black Hat, USA, 2004.
35. O. Yavanoglu and M. Aydos, "A review on cyber security datasets for machine learning algorithms," in Proceedings of the IEEE International Conference on Big Data (Big Data) 2017, pp. 2186-2193, doi: 10.1109/BigData.2017.8258167.
36. '2019 Cost of a Data Breach Study: Global Overview'. Ponemon Institute and IBM, 2019. Accessed: October 2019. [Online]. Available: https://www.ibm.com/security/data-breach
37. "Precision and recall," Wikipedia, Accessed: October 22, 2019. [Online]. Available: https://en.wikipedia.org/wiki/Precision_and_recall
38. "Shift Plan," Wikipedia, Accessed: November 18, 2019. [Online]. Available: https://en.wikipedia.org/wiki/Shift_plan.